\pdfoutput=1
\documentclass{article}
\usepackage{frascatiphys,graphicx,subfigure}
\usepackage[latin1]{inputenc}
\usepackage{cite}
\usepackage{listings} 
\newcommand{\beq}{\begin{equation}}
\newcommand{\eeq}{\end{equation}}
\newcommand{\bea}{\begin{eqnarray}}
\newcommand{\eea}{\end{eqnarray}}
\newcommand{\real}{{\sf I}\kern-.12em{\sf R}}
\newcommand{\comp}{{\sf I}\kern-.50em{\sf C}}
\newcommand{\unity}{{\sf I}\kern-.54em{\sf 1}}

\newcommand{\vsb}{\vspace{-0.16cm}}
\newcommand{\stringa}{\ttfamily\lstinline}
\def\cod#1{{\stringa!#1!}}
\begin{document}
\title{\textbf{BFKL phenomenology: resummation 
               of high-energy logs 
               in semi-hard processes at LHC}}
\author{
Francesco Giovanni Celiberto \\
{\em Dipartimento di Fisica, Università della Calabria}, 
\\
{\em Istituto Nazionale di Fisica Nucleare, 
     Gruppo Collegato di Cosenza} 
\\
{\em Arcavacata di Rende, 87036 Cosenza, Italy}
}
\maketitle
\baselineskip=11.6pt
\begin{abstract}
A study of differential cross sections 
and azimuthal observables for semi-hard processes 
at LHC energies, including BFKL resummation effects, 
is presented. Particular attention 
has been paid to the behaviour 
of the azimuthal correlation momenta, 
when a couple of forward/backward jets 
or identified hadrons is produced in the final state 
with a large rapidity separation. 
Three- and four- jet production has been also considered, 
the main focus lying on the definition of new, 
generalized azimuthal observables, whose dependence 
on the transverse momenta and the rapidities 
of the central jet(s) can be considered 
as a distinct signal of the onset of BFKL dynamics.
\end{abstract}
\section{Introduction}
The large amount of data already recorded 
and to be produced
in the near future at the Large Hadron Collider (LHC) 
offers a peerless opportunity to probe 
perturbative QCD at high energies.
Multi-Regge kinematics (MRK), which prescribes the production 
of strongly rapidity-ordered objects in the final state,
is the key point for the study of semi-hard processes 
in the high-energy limit. In this kinematical regime, 
the Balitsky-Fadin-Kuraev-Lipatov 
(BFKL) approach, at leading (LL) 
\cite{Lipatov:1985uk,Balitsky:1978ic,
Kuraev:1977fs,Kuraev:1976ge,Lipatov:1976zz,Fadin:1975cb} 
and next-to-leading 
(NLL)~\cite{Fadin:1998py,Ciafaloni:1998gs} accuracy, 
represents perhaps the most powerful tool
to perform the resummation of large logarithms 
in the colliding energy 
to all orders of the perturbative expansion.
So far, Mueller--Navelet jet production~\cite{Mueller:1986ey} 
has been the most studied reaction.
Interesting observables associated to this process are 
the azimuthal correlation momenta, which, however, 
are strongly affected by collinear contaminations.
Therefore, new observables, independent 
from these contaminations, 
were proposed in~\cite{Vera:2006un,Vera:2007kn} 
and calculated at NLL 
in~\cite{Colferai2010,Ducloue2013,Ducloue2014,Ducloue:2014koa,
Caporale:2014gpa,Celiberto:2015yba,Celiberto:2016ygs,
Ciesielski:2014dfa,Angioni:2011wj,Chachamis:2015crx},  
showing a very good agreement 
with experimental data at the LHC.
Unfortunately, Mueller--Navelet configurations 
are still too inclusive to perform MRK precision studies. 
With the aim to further and deeply probe the BFKL dynamics, 
we propose to investigate two different kinds of processes.
The first one is the detection of two charged light hadrons:
$\pi^{\pm}$, $K^{\pm}$, $p$, $\bar p$ 
having high transverse momenta and separated by a large interval 
of rapidity, together with an undetected hadronic
system $X$~\cite{Ivanov:2012iv,Celiberto:2016hae}.
On one side, hadrons can be detected at the LHC 
at much smaller values of the transverse
momentum than jets, allowing us to explore 
a kinematic range outside 
the reach of the Mueller--Navelet channel. 
On the other side, this process makes it possible to constrain
not only the parton densities (PDFs) 
for the initial proton, but also 
the parton fragmentation functions (FFs) 
describing the detected hadron in the final state. 
The second kind of processes 
is the multi-jet production~\cite{Caporale:2015int,
Caporale:2015vya,Caporale:2016soq,Caporale:2016xku}, 
which allows to define new, generalized and suitable 
BFKL observables by considering extra jets well separated 
in rapidity in the final state and by studying the dependence 
on their transverse momenta and azimuthal angles.

\section{Di-hadron production}

\begin{figure}[h]
\includegraphics[scale=0.225]{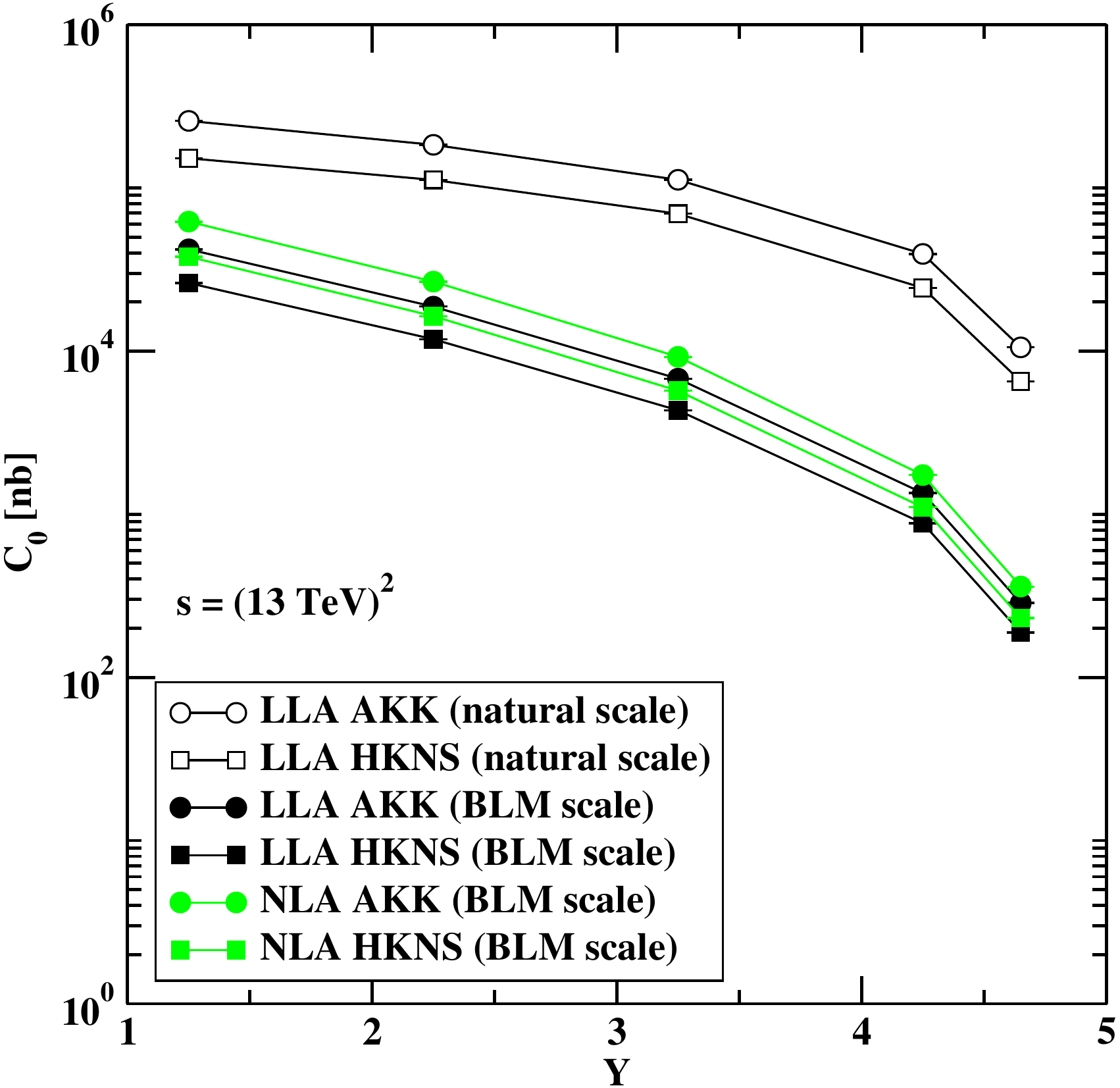}
\includegraphics[scale=0.225]{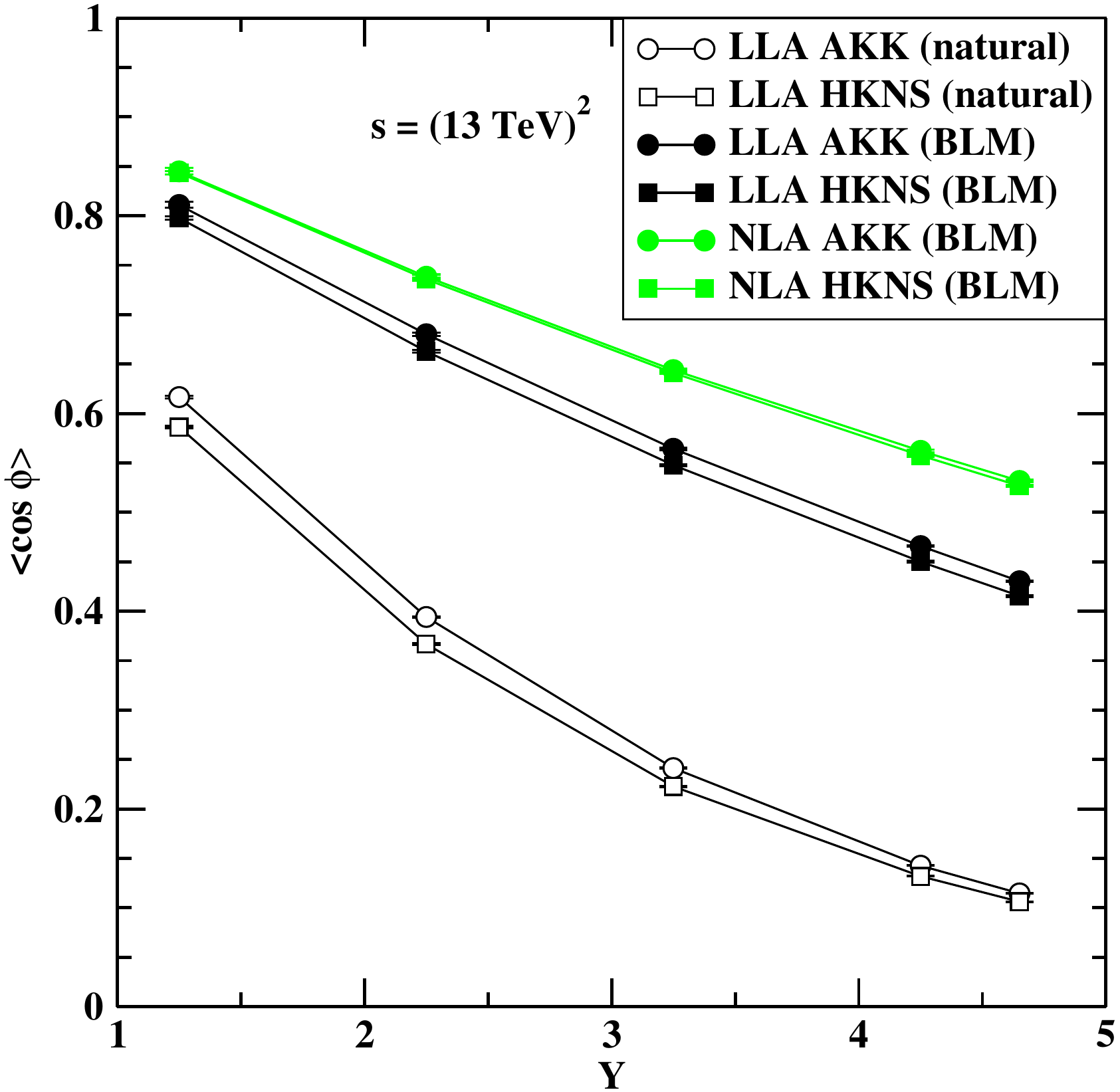}
\includegraphics[scale=0.225]{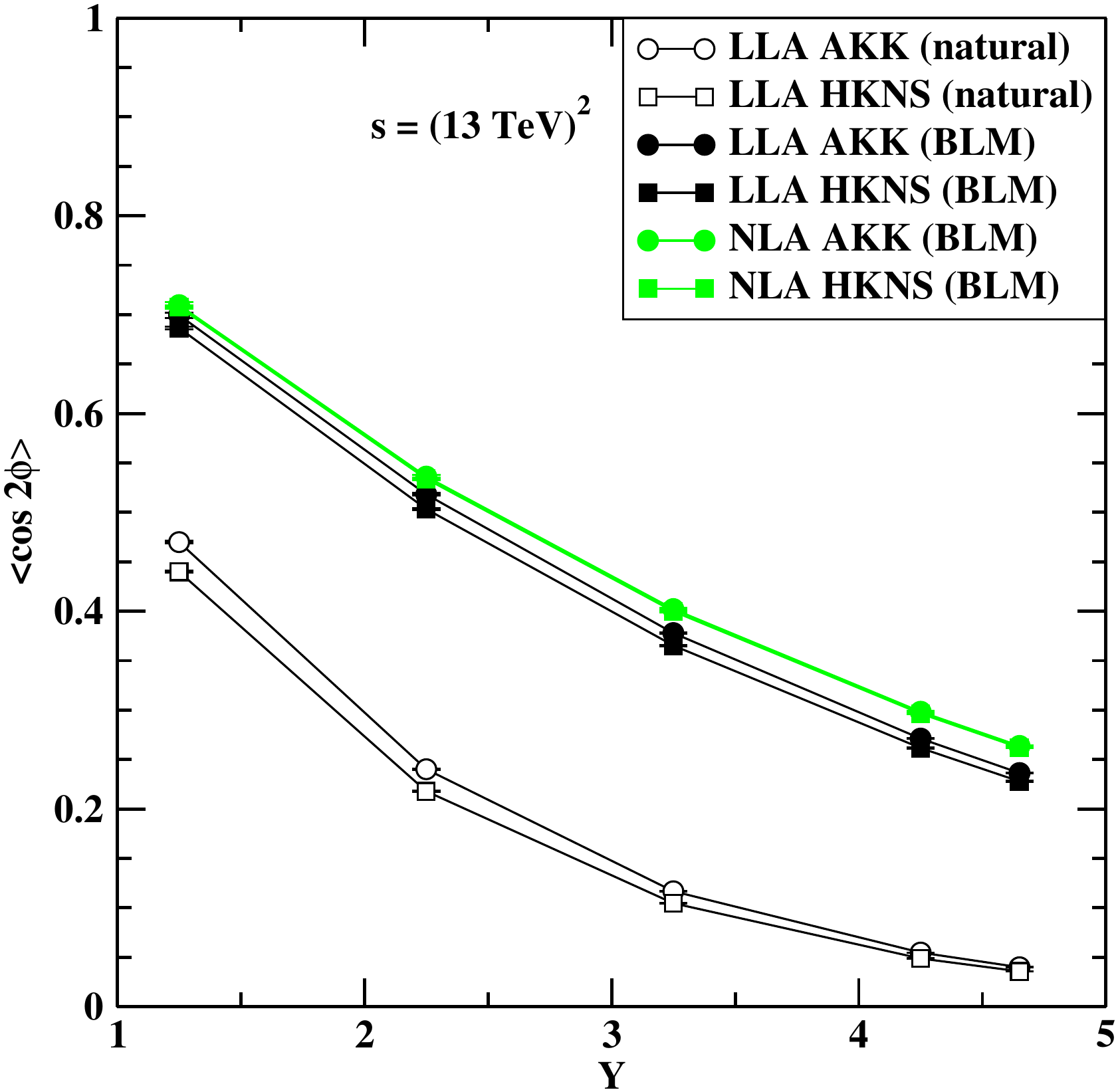}
\caption{$Y$ dependence of cross section, 
$\langle \cos \phi \rangle$
and $\langle \cos 2\phi \rangle$ 
for di-hadron production at $\sqrt{s}=13$ TeV. 
See Ref.~\cite{Celiberto:2016hae} 
for the FF parametrizazions used 
and for the definition 
of ``natural'' and ``BLM'' scales.}
\label{fig:hadrons}
\end{figure}

We consider the production, in high-energy 
proton-proton collisions, of a pair 
of identified hadrons with large transverse momenta, 
$\vec k_1^2\sim \vec k_2^2 \gg \Lambda^2_{\rm QCD}$ 
and large separation in rapidity.
The differential cross section of the process reads  
\beq
\frac{d\sigma^{\rm di-hadron}}{dy_1dy_2\, d|\vec k_1| \, d|\vec k_2|d\phi_1 d\phi_2}
=
\frac{1}{(2\pi)^2}
\left[
{\cal C}_0+\sum_{n=1}^\infty  2\cos (n\phi ){\cal C}_n \right] \;,
\eeq
where $\phi=\phi_1-\phi_2-\pi$, with $\phi_{1,2}$ the two hadrons'
azimuthal angles, while $y_{1,2}$ and $\vec k_{1,2}$ are their
rapidities and transverse momenta, respectively. 
In order to match the kinematic cuts 
used by the CMS collaboration, we consider 
the \emph{integrated azimuthal coefficients}
given by
\begin{equation}\label{Cm_int}
C_n=
\int_{y_{1,\rm min}}^{y_{1,\rm max}}\hspace{-0.21cm}dy_1
\int_{y_{2,\rm min}}^{y_{2,\rm max}}\hspace{-0.21cm}dy_2
\int_{k_{1,\rm min}}^{\infty}\hspace{-0.21cm}dk_1
\int_{k_{2,\rm min}}^{\infty}\hspace{-0.21cm}dk_2
\delta\left(y_1-y_2-Y\right)
{\cal C}_n 
\end{equation}
and their ratios $R_{nm}\equiv C_n/C_m$.
For the integrations over rapidities and transverse momenta
we use the limits, 
$y_{1,\rm min}=-y_{2,\rm max}=-2.4$, 
$y_{1,\rm max}=-y_{2,\rm min}=2.4$, 
$k_{1,\rm min}=k_{2,\rm min}=5$ GeV,
which are realistic values 
for the identified hadron detection at LHC.
In Fig.~\ref{fig:hadrons} the dependence on the
rapidity separation between the detected hadrons, 
$Y=y_1-y_2$, of the $\phi$-averaged cross section 
$C_0$ and of the ratios $R_{10}$ and $R_{20}$ 
at the center-of-mass energy $\sqrt{s} = 13$ TeV is shown.

\section{Multi-jet production}

\begin{figure}[t]
\centering
\includegraphics[scale=0.29]{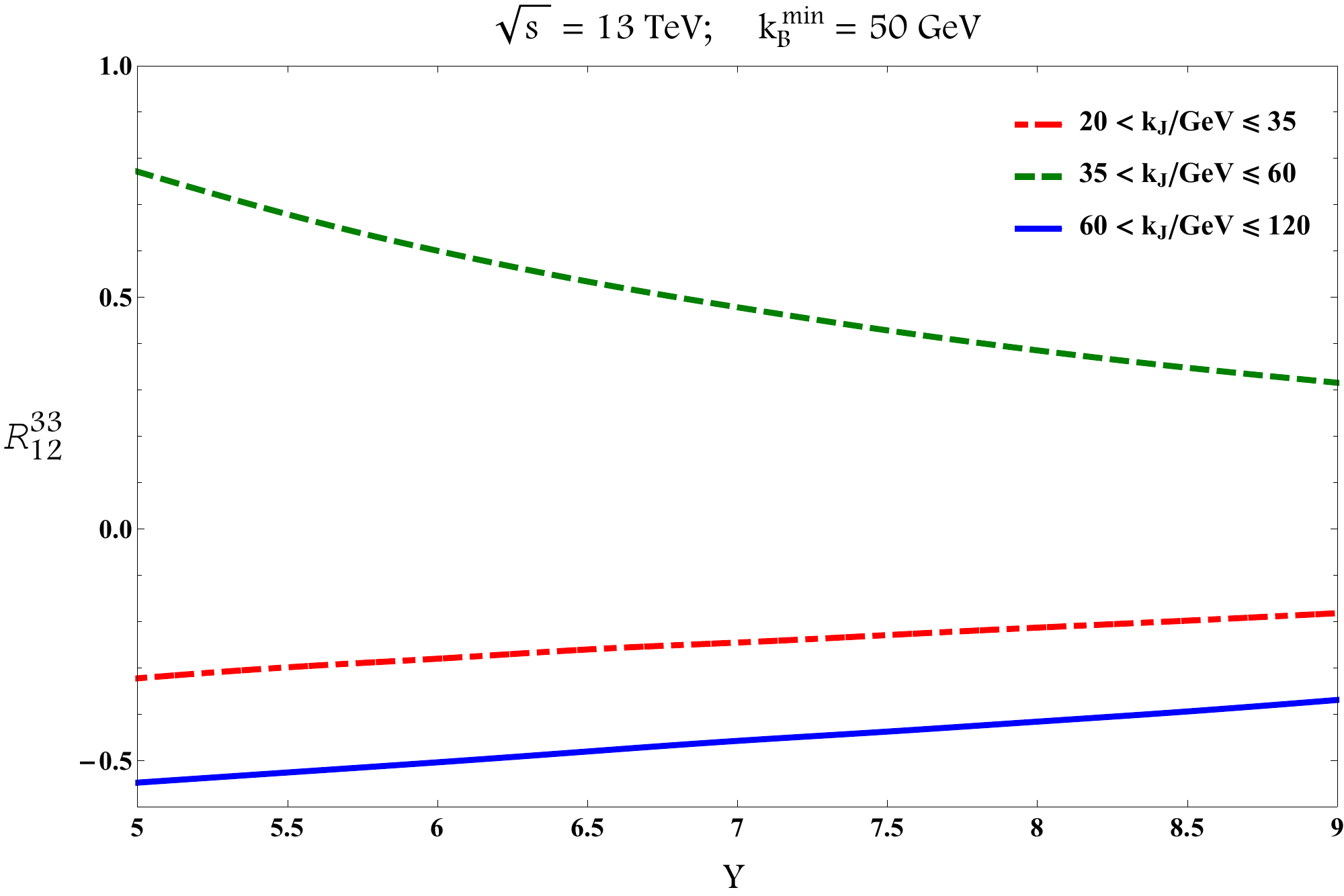}
\includegraphics[scale=0.29]{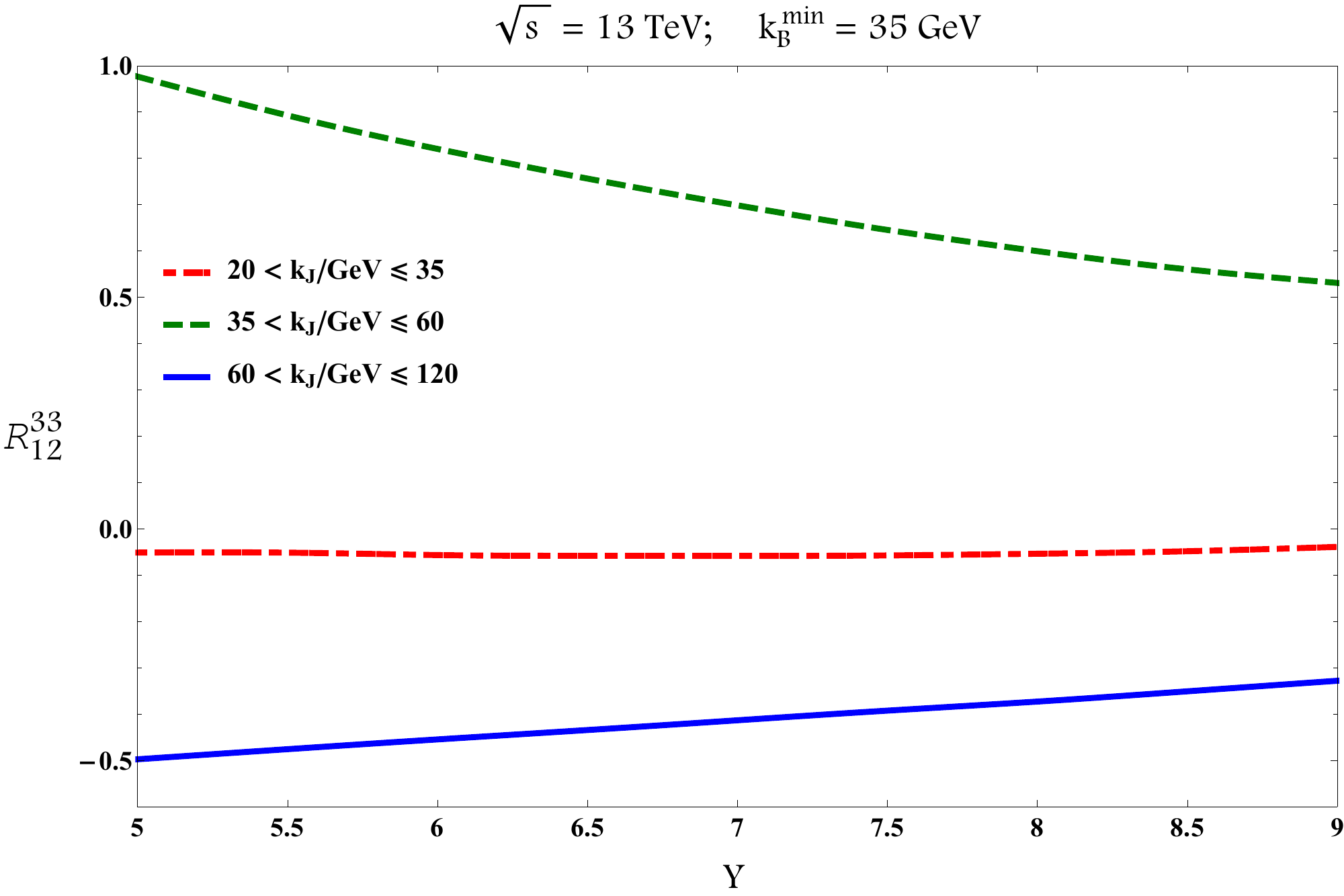}
\caption{$Y$dependence of $R^{33}_{12}$ for $\sqrt{s} = 13$ TeV 
and $k_{B,\rm min}$ = 50 GeV (left column)
and $k_{B,\rm min}$ = 35 GeV (right column). 
$k_{A,\rm min}$ is fixed to 35 GeV, 
while the central jet has rapidity $y_J = (y_A + y_B)/2$.}
\label{fig:3-jet}
\end{figure}

\begin{figure}[t]
\centering
\includegraphics[scale=0.165]{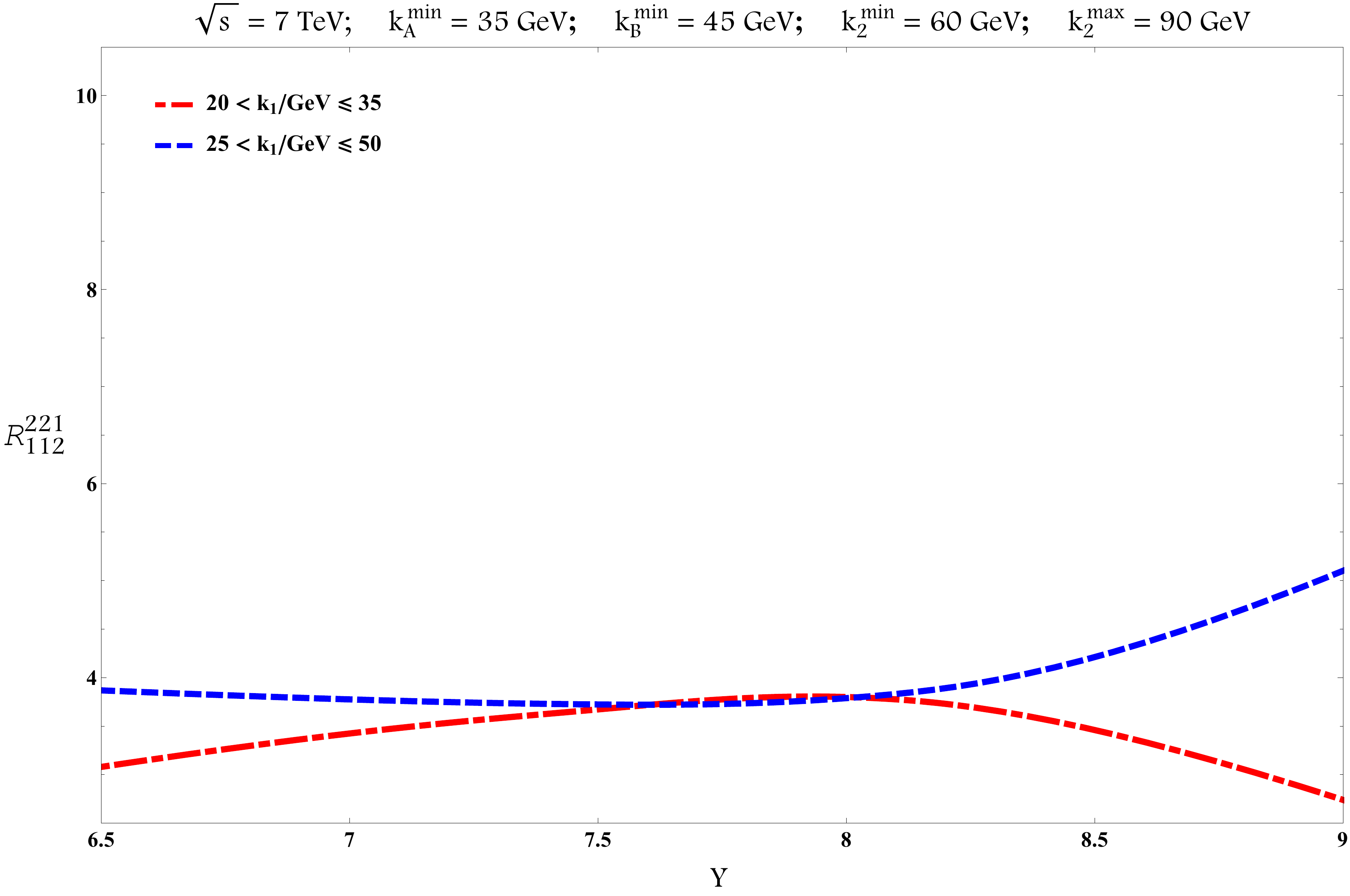}
\includegraphics[scale=0.165]{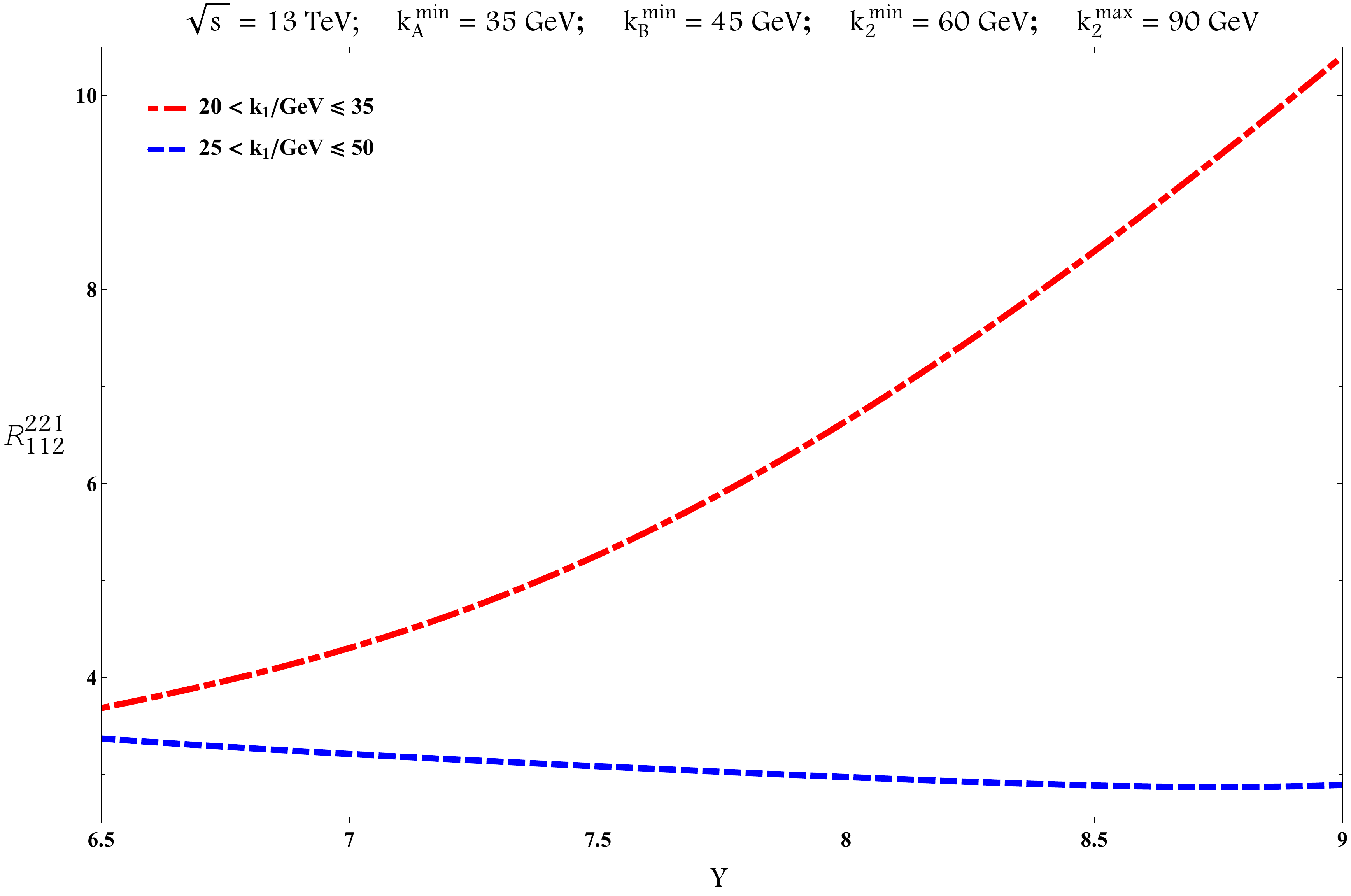}
\caption{$Y$dependence of $R^{221}_{112}$ 
for $\sqrt{s} = 7$ TeV (left column)
and for $\sqrt{s} = 13$ TeV (right column). 
The rapidity interval between a jet and the closest one 
is fixed to $Y/3$.}
\label{fig:4-jet}
\end{figure}

The process under investigation 
is the hadroproduction of $n$ jets in the final state, 
well separated in rapidity so that 
$y_i > y_{i+1}$ according to MRK, 
and with their transverse momenta $\{k_i\}$ 
lying above the experimental resolution scale,
together with an undetected soft-gluon radiaton emission.
Pursuing the goal to generalize the azimuthal ratios $R_{nm}$ 
defined for Mueller--Navelet jet and di-hadron production, 
we define new, generalized azimuthal correlation momenta 
by projecting the differential cross section 
$d\sigma^{n-{\rm jet}}$
on all angles, so having 
\begin{equation}
\mathcal{C}_{M_1 \cdots M_{n-1}} =
\left\langle 
 \prod_{i=1}^{n-1} \cos\left(M_i \, \phi_{i,i+1}\right)
\right\rangle = \hspace{-0.1cm} 
\int_0^{2\pi} \hspace{-0.4cm} d\theta_1 
\cdots \hspace{-0.1cm} 
\int_0^{2\pi} \hspace{-0.4cm} d\theta_n
\prod_{i=1}^{n-1} \cos\left(M_i \, \phi_{i,i+1}\right)
d\sigma^{n-{\rm jet}}
\end{equation}
where $\phi_{i,i+1} = \theta_i - \theta_{i+1} - \pi$, 
with $\theta_i$ being the azimuthal angle of the jet $i$.
Firstly, we introduce realistic LHC kinematical cuts
by integrating $\mathcal{C}_{M_1 \cdots M_{n-1}}$ 
over rapidities and momenta of the tagged jets
\begin{equation}\label{Cm_int}
C_{M_1 \cdots M_{n-1}} =
\int_{y_{1,\rm min}}^{y_{1,\rm max}}\hspace{-0.5cm}dy_1
\cdots
\int_{y_{n,\rm min}}^{y_{n,\rm max}}\hspace{-0.5cm}dy_n
\int_{k_{1,\rm min}}^{\infty}\hspace{-0.5cm}dk_1
\cdots
\int_{k_{n,\rm min}}^{\infty}\hspace{-0.5cm}dk_n
\delta\left(y_1-y_n-Y\right)
{\cal C}_n 
\end{equation}
and by keeping fixed the rapidity difference $Y = y_1 - y_n$ 
between the most forward and the most backward jet, which 
corresponds to the maximum rapidity interval in the final state.
Secondly, we remove the zeroth conformal spin contribution 
responsible for any collinear contamination 
and we minimise possible higher-order effects 
by studying the ratios 
$R^{M_1 \cdots M_{n-1}}_{N_1 \cdots N_{n-1}} \equiv 
C_{M_1 \cdots M_{n-1}}/C_{N_1 \cdots N_{n-1}}$ 
where $\{M_i\}$ and $\{N_i\}$ are positive integers.
In Fig.~\ref{fig:3-jet} we show the dependence on $Y$ 
of the coefficient $R^{33}_{12}$, 
characteristic of the 3-jet production process, 
for $\sqrt{s} = 13$ TeV, for two different kinematical cuts 
on the transverse momenta $k_{A,B}$ of the external jets 
and for three different ranges of the central jet 
transverse momentum $k_J$.
In Fig.~\ref{fig:4-jet} we show the dependence on $Y$ 
of the coefficient $R^{221}_{112}$, 
characteristic of the 4-jet production process, 
for $\sqrt{s} = 7$ and $13$ TeV, for asymmetrical cuts 
on the transverse momenta $k_{A,B}$ of the external jets
and for two different configurations 
of the central jet transverse momenta $k_{1,2}$. 
A comparison with predictions for these observables from
fixed order analyses as well as from the BFKL inspired 
Monte Carlo {\bf\cod{BFKLex}}~\cite{Chachamis:2011rw,
Chachamis:2011nz,Chachamis:2012fk,
Chachamis:2012qw,Chachamis:2015ico} 
is underway.

\section{Conclusions}
We perfomed a study of perturbative QCD 
in the high-energy limit through 
two different classes of processes.
First we investigated the behaviour of cross section 
and azimuthal ratios for di-hadron production, 
which represents a less inclusive final state process 
with respect to the well known Mueller--Navelet jet reaction.
Then we proposed to study multi-jet production processes, 
in order to define new, 
generalized and suitable BFKL observables.
The comparison with experimental data will help to gauge 
and disentangle the applicability region of the BFKL formalism, 
therefore it is needed and suggested.



%
\end{document}